\documentclass[aps,prmaterials,reprint,groupedaddress]{revtex4-2}

\usepackage{graphicx}
\usepackage[euler]{textgreek}
\usepackage[english]{babel}

\begin{document}

\title{Ionic liquid dynamics in nanoporous carbon: A pore-size- and temperature-dependent neutron spectroscopy study on supercapacitor materials}

\author{Mark Busch}
\email{mark.busch@tuhh.de}
\affiliation{Institute of Materials Physics and Technology, Hamburg University of Technology, Ei{\ss}endorfer Stra{\ss}e 42, 21073 Hamburg, Germany}

\author{Tommy Hofmann}
\affiliation{Helmholtz-Zentrum Berlin f{\"u}r Materialien und Energie, Hahn-Meitner-Platz 1, 14109 Berlin, Germany}

\author{Bernhard Frick}
\affiliation{Institut Laue-Langevin, 71 Avenue des Martyrs, 38000 Grenoble, France}

\author{Jan P. Embs}
\affiliation{Laboratory for Neutron Scattering and Imaging, Paul Scherrer Institute, 5232 Villigen, Switzerland}

\author{Boris Dyatkin}
\affiliation{Department of Materials Science and Engineering, A. J. Drexel Nanomaterials Institute, Drexel University, Philadelphia, PA, 19104, U.S.A.}

\author{Patrick Huber}
\email{patrick.huber@tuhh.de}
\affiliation{Institute of Materials Physics and Technology, Hamburg University of Technology, Ei{\ss}endorfer Stra{\ss}e 42, 21073 Hamburg, Germany}

\date{\today}

\begin{abstract}
	The influence of spatial confinement on the thermally excited stochastic cation dynamics of the room-temperature ionic liquid 1-N-butylpyridinium bis-((trifluoromethyl)sulfonyl)imide ([BuPy][Tf$_2$N]) inside porous carbide-derived carbons with various pore sizes in the sub- to a few nanometer range are investigated by quasi-elastic neutron spectroscopy. Using the potential of fixed window scans, i.e.\ scanning a sample parameter, while observing solely one specific energy transfer value, an overview of the dynamic landscape within a wide temperature range is obtained. It is shown that already these data provide a quite comprehensive understanding of the confinement-induced alteration of the molecular mobility in comparison to the bulk. A complementary, more detailed analysis of full energy transfer spectra at selected temperatures reveals two translational diffusive processes on different time scales. Both are considerably slower than in the bulk liquid and show a decrease of the respective self-diffusion coefficients with decreasing nanopore size. Different thermal activation energies for molecular self-diffusion in nanoporous carbons with similar pore size indicate the importance of pore morphology on the molecular mobility, beyond the pure degree of confinement. In spite of the dynamic slowing down we can show that the temperature range of the liquid state upon nanoconfinement is remarkably extended to much lower temperatures, which is beneficial for potential technical applications of such systems.
\end{abstract}

\maketitle

\section{Introduction}

Room-temperature ionic liquids electrolytes in combination with porous carbon electrodes are promising candidates for electric double layer capacitors (EDLCs), which are commonly referred to as super- or ultracapacitors.\cite{Simon2008,Zhai2011,Zhang2017,Osti2020,Huber2020} These capacitors benefit from high energy densities on a par with those of batteries along with the superior power densities of conventional capacitors and thus offer efficient electric energy storage and conversion systems for a variety of applications.
The use of ionic liquids has numerous advantages as compared to conventional aqueous or solvated organic salt electrolytes. Most importantly, they possess a wide electrochemical window, good temperature stability and low volatility.\cite{Simon2008, Armand2009, Zhang2017, Vatamanu2017, Salanne2017} Nanoporous carbons distinguish themselves by a well-tunable pore size, shape and surface chemistry. Furthermore, they possess a high electrical conductivity, good \mbox{(electro-)} chemical stability and a large specific surface area, making them well-suited electrode materials.\cite{Cherusseri2019}
Although, these systems are a very active field of research, the actual technical implementation as high-performance supercapacitors remains challenging. One of the reasons is the relatively high viscosity of ionic liquids \cite{Zhang2006a, Nazet2017, Salanne2017}, resulting in slow ion dynamics.\cite{Panesar2013, Salanne2017} While some studies find increased dynamics in the case of carbon nanoconfinement \cite{Chathoth2012, Chathoth2013, Chaban2014, Berrod2016}, others see a clear slowdown.\cite{Singh2010a, Li2013, Dyatkin2016, Dyatkin2018} Additionally, ionic liquids tend to layer at solid interfaces \cite{Mezger2008, Griffin2017}. As a consequence, they often form immobile layers at pore walls and molecules within micropores thus frequently do not exhibit any appreciable diffusive motion \cite{Chathoth2012, Chathoth2013, Banuelos2014, Dyatkin2018, Dyatkin2018a}, while even confinement-induced freezing has been reported \cite{Comtet2017}. This further reduces the amount of mobile ions in the nanopores and thus potentially decreases the supercapacitor performance, since (dis-)charging involves diffusive processes.\cite{Kondrat2014}\\
To resolve these uncertainties, more profound knowledge about the self-diffusion properties of ionic liquids in carbon nanoconfinement must be obtained. In this context, the influence of different nanopore sizes on the dynamics is not only of importance with respect to possible immobilised ionic surface layers, but also because of the dependence of the capacitance on the pore diameter. It has been found that the capacitance of ionic liquid based electrical double layer capacitors depends not monotonically, but rather in an oscillatory manner, on the pore size.\cite{Feng2011, Jiang2011, Wu2011} Thus, additional research efforts must find the optimal nanopore width that yields the best compromise between capacitance and molecular mobility of the ionic liquid molecules. The latter, however, usually decreases with lower temperatures and the temperature range of the liquid phase is often quite limited for bulk ionic liquids, especially towards low temperatures.\cite{Chathoth2012, Zhang2017, Tokuda2006a, Zhang2006a, Salanne2017} But nanoconfinement of ionic liquids is known to change the phase transition behaviour.\cite{Kanakubo2006, Weingarth2014}\\
To address these fundamental issues, we present a quasi-elastic neutron scattering investigation on the self-diffusion dynamics of a room-temperature ionic liquid confined in the nanopores of carbide-derived carbons (CDCs) as a function of pore size over a wide temperature range.\\
In a first step, we analyse the data of so-called fixed window scans, which allow to rapidly scan a broad temperature range and provide a first, but already detailed overview of the dynamics in our systems. In a second step, these findings are complemented at selected temperatures by the full spectroscopic information, from two spectrometers with complementary energy resolution and dynamic range, providing further insights into certain dynamic peculiarities of the nanoconfined ionic liquid.

\section{Sample preparation \& characterisation}

\begin{figure}
	\vspace*{4mm}
	\includegraphics{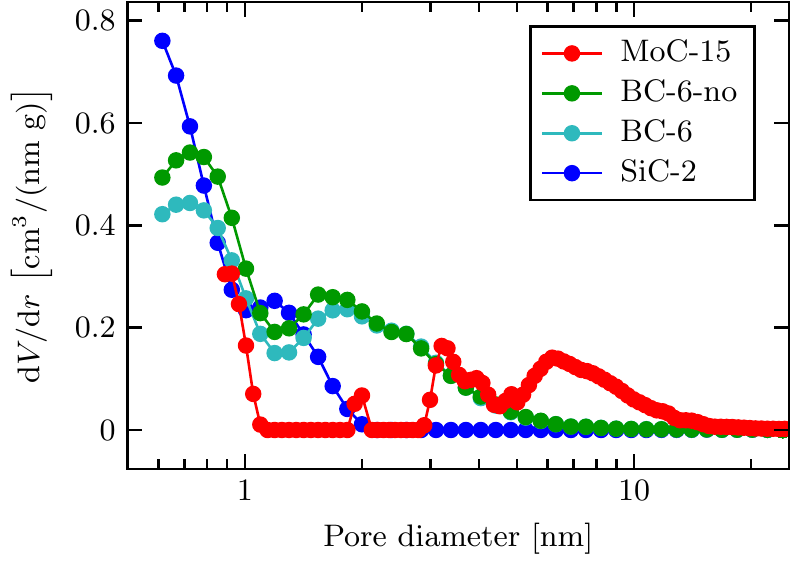}
	\caption{Pore size distribution of the carbide-derived nanoporous carbon samples, as obtained using a density functional theory based analysis of the nitrogen sorption isotherms (Lines between the points are guides for the eye.).}
	\label{fig:PSD}
\end{figure}

\begin{figure}
	\includegraphics[width=\columnwidth]{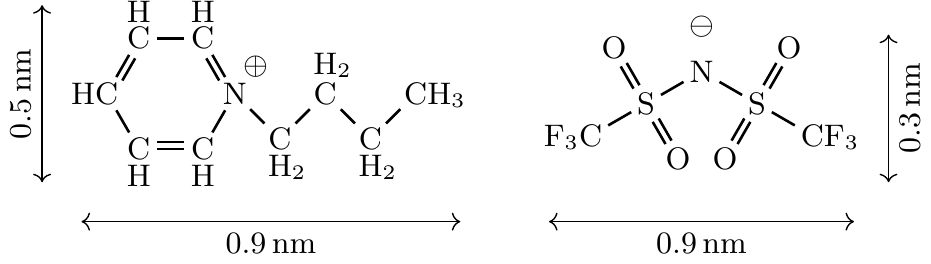}
	\caption{Structure formula of the cation (\textit{left}) and the anion (\textit{right}) of the ionic liquid {[BuPy]}{[Tf$_2$N]} with estimate of the ions' size.}
	\label{fig:BuPyTf2N_formula}
\end{figure}

Porous CDC microparticles were obtained using previously reported procedures.\cite{Presser2011} Silicon carbide (SiC), molybdenum carbide (Mo$_2$C) and boron carbide (B$_4$C) microparticles \mbox{(1-5\,\textmu m} wide) were used as precursors. These were etched with Cl$_2$ gas at specific temperatures to remove the metal atoms and structure sample-specific nanoporosity and were subsequently annealed in H$_2$ gas at 600\,$^\circ$C. SiC was etched at 1000\,$^\circ$C with Cl$_2$ and following the H$_2$ annealing, it was oxidised in air at 425\,$^\circ$C and annealed under high vacuum (10$^{-6}$ torr) at 1400\,$^\circ$C (labelled as 'SiC-2'). Mo$_2$C was etched at 900\,$^\circ$C with Cl$_2$ and after the H$_2$ annealing it underwent identical oxidation and annealing conditions as in the case of SiC (labelled as 'MoC-15'). After the etching of B$_4$C with Cl$_2$ at 900\,$^\circ$C and the following H$_2$ annealing, it underwent either the same oxidation and annealing conditions as above (labelled as 'BC-6') or vacuum annealing only, without preceding air oxidation step (labelled as 'BC-6-no'). While the vacuum annealing final step removes all surface functional groups and makes electrode surface chemistries of each system near-identical \cite{Dyatkin2014, Dyatkin2015}, the preceding steps yielded different pore widths.\\
Figure~\ref{fig:PSD} shows the pore size distribution of the CDCs, as obtained from the nitrogen sorption isotherms, employing quenched solid density functional theory models.\cite{Neimark2009}
The numbers at the end of the sample identifiers indicate the largest nanometre-sized pore diameter determined for the respective specimen.\\
Subsequently the nanopores of the carbide-derived carbons are filled with a room-temperature ionic liquid 1-N-butylpyridinium bis-((trifluoromethyl)sulfonyl)imide, in the following shortly denoted as [BuPy][Tf$_2$N] and whose bulk molecular dynamics have been already extensively studied.\cite{Embs2012, Embs2013, Burankova2014, Burankova2014b, Burankova2017} The structure formulas of the cation and anion are depicted in Fig.~\ref{fig:BuPyTf2N_formula}.
\begin{figure*}
	\vspace*{4mm}
	\includegraphics{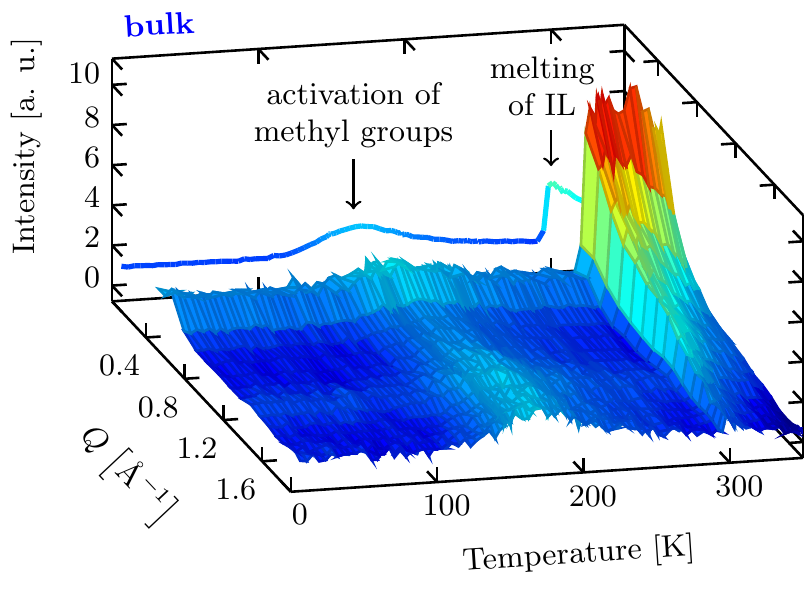}
	\hspace*{10mm}
	\includegraphics{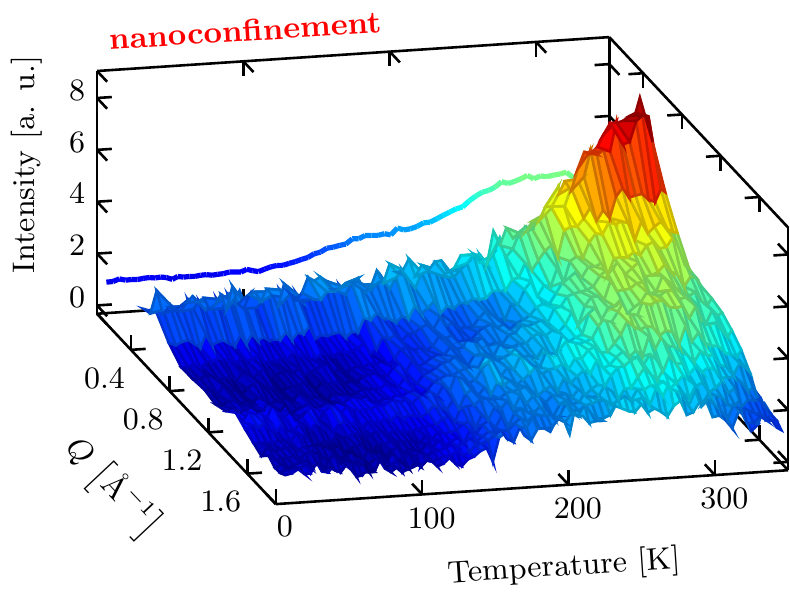}
	\caption{Intensity of neutrons with an energy transfer of $\pm 2\, \text{\textmu eV}$, when scattered at \textit{(left)} bulk {[BuPy]}{[Tf$_2$N]} and \textit{(right)} in the nanoconfinement of MoC-15 as a function of wave vector transfer $Q$, while scanning the temperature from 2\,K to 355\,K with a rate of 1\,K/min. \textit{Back panel:} Intensity averaged over the available $Q$-range.}
	\label{fig:IFWS_bulkIL_3D}
	\label{fig:IFWS_S4_3D}
\end{figure*}
Filling of the nanoporous carbons with the ionic liquid for the neutron spectroscopy experiments is a crucial process. Since the aim is to study the dynamics of the ionic liquid confined in the nanopores, any bulk liquid between the carbon grains must be avoided, as that could overwhelm the measurement signal from the confined liquid. Therefore, for each sample, only an amount of ionic liquid equalling the pore volume obtained from the nitrogen gas sorption data analysis is used. That volume of [BuPy][Tf$_2$N] is mixed with ethanol ($\geq$99.9\,\% purity) in a volume ratio of 1:2 and afterwards blended with the respective CDC. The compound is stored in a drying cabinet for 2.5 hours at 75\,$^\circ$C and 8\,mbar, to remove the ethanol and accelerate the capillary imbibition of the highly viscous ionic liquid into the nanopores. Pore filling of the now visually dry powder is verified gravimetrically.

\section{Methods}

Quasi-elastic neutron scattering (QENS) experiments have been performed at the time-of-flight spectrometer \mbox{FOCUS} at the Paul Scherrer Institute (Villigen, Switzerland) and the neutron backscattering spectrometer \mbox{IN16B} at the Institut Laue-Langevin (Grenoble, France).\cite{Busch2015x}
At \mbox{FOCUS} neutrons with a wavelength of 6.00\,\AA\ were used, giving access to an energy transfer range of $\pm$1.0\,meV with a simultaneously accessible wave vector transfer $Q$ between 0.32\,\AA$^{-1}$ and 1.64\,\AA$^{-1}$ and an energy resolution of around {39\,\textmu eV} (FWHM).
\mbox{IN16B} was used in its low-background position, with an incident neutron wavelength of 6.27\,\AA. Here, the maximum energy transfer $\Delta E$ is $\pm$31\,{\textmu eV}, with resolution of around {0.85\,\textmu eV} (FWHM) and a Q-range from 0.19\,\AA$^{-1}$ to 1.90\,\AA$^{-1}$.
At \mbox{IN16B} not only full quasi-elastic spectra at selected temperatures were acquired, but also so-called fixed windows scans were performed. The latter enables one to quickly scan a very broad temperature range, while observing the intensity of a fixed $\Delta E$ channel of the spectrum.\cite{Frick2012} In case of the full quasi-elastic spectra data from an empty cell measurement was subtracted from the specimen data.\\
The incoherent neutron scattering cross section of the cation of the ionic liquid is 1124.24\,barn, while the one of the anion is with only 0.53\,barn considerably smaller.\cite{Sears1992} As a result it is mainly the dynamics of the cation that is probed by the inelastic neutron scattering methods employed in this work. As for the type of cation dynamics we can expect contributions from local motions (ring wagging, segmental rotation in the butyl side chain, methyl group rotation) and from centre-of-mass diffusion. In general local motions are characterised in QENS by a $Q$-independence of the line broadening, whereas diffusive processes show a pronounced $Q$-dependence.\\
At both spectrometers the sample holder was placed in a cryostat enabling a temperature control of the specimens.
All samples were encased in flat, slab-shaped aluminium sample holders.
The primary data reduction, detector efficiency calibration utilising a vanadium standard and corrections like those concerning self-shielding were done using the \mbox{DAVE} \cite{Azuah2009} \mbox{(FOCUS)} and the \mbox{LAMP} \cite{Richard1996} \mbox{(IN16B)} software packages.
In principal after these corrections the data can be converted into the dynamic structure factor $S(Q,\omega)$ which can then be compared to theoretical models after convolution with the instrumental resolution.

\section{Experimental results \& discussion}

\subsection{Fixed window scans}

\begin{figure}
	\vspace*{4mm}
	\includegraphics{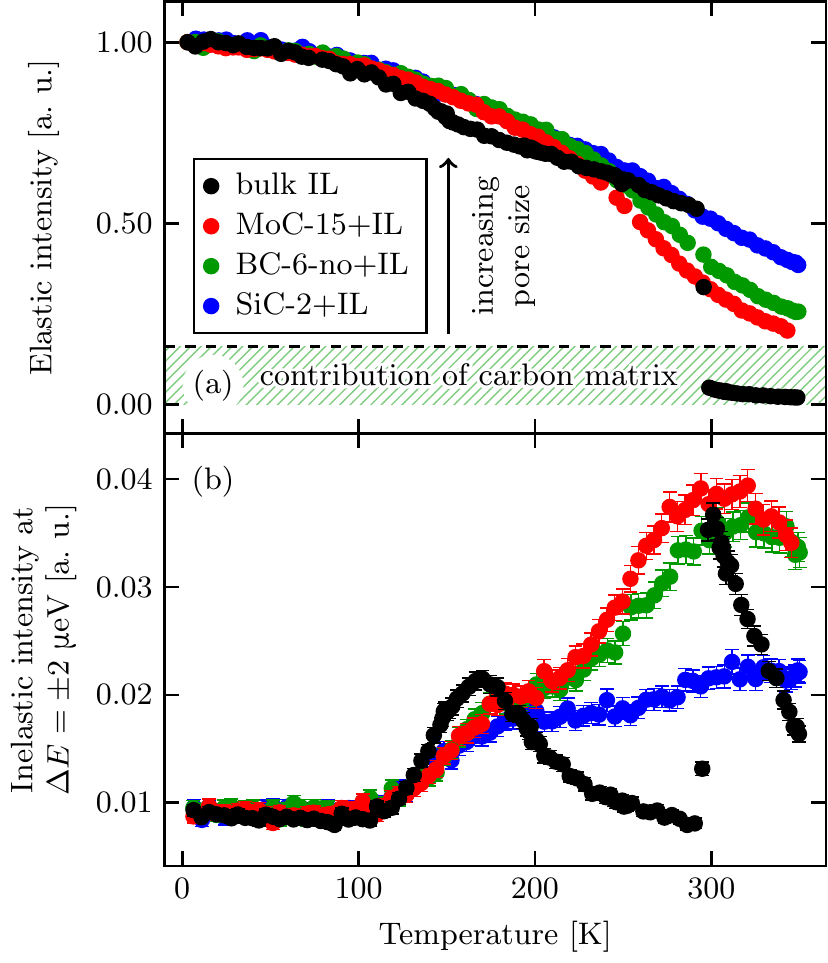}
	\caption{Fixed window scans in a temperature range from 2\,K to 355\,K with a heating rate of 1\,K/min for the ionic liquid in bulk and in nanoconfinement of different pore sizes (Intensity averaged over a $Q$ range from 0.44\,\AA$^{-1}$ to 1.90\,\AA$^{-1}$): (a) Elastic fixed window scans, i.e.\ $\Delta E= 0\,\text{\textmu eV}$, (b) Inelastic fixed window scans at an energy transfer of $\Delta E = \pm 2\,\text{\textmu eV}$.}
	\label{fig:FWS_1D}
\end{figure}

To get an overview of the various dynamic processes setting in at different temperatures in the systems under investigation, elastic (EFWS) and inelastic fixed window scans (IFWS) are conducted in a temperature range from 2\,K to 355\,K. While heating the specimens with a rate of 1\,K/min, alternately the intensity of the elastically scattered neutrons, i.e.\ $\Delta E = 0\,\text{\textmu eV}$ and those having experienced an energy change of $\Delta E = \pm 2\,\text{\textmu eV}$ is measured.
Figure~\ref{fig:IFWS_bulkIL_3D} shows the acquired data from such an IFWS at the example of the bulk [BuPy][Tf$_2$N] (left panel), which is measured as a reference for the subsequent analysis of the respective cation dynamics in nanoconfinement (right panel). This comparison already impressively illustrates the confinement-induced alterations in the phase transition behaviour, as well as the $Q$-dependence of the data at different temperatures, giving first hints on the nature of the molecular dynamics.
A further qualitative overview of the temperature-dependent activation of dynamic processes is given by the averaged intensity over the $Q$ range between 0.44\,\AA$^{-1}$ and 1.90\,\AA$^{-1}$, that is depicted in the back panel of Fig.~\ref{fig:IFWS_bulkIL_3D}, as well as in Fig.~\ref{fig:FWS_1D} for the bulk [BuPy][Tf$_2$N] and confined inside the nanopores of the CDCs with different pore sizes.
Here, all intensities, i.e.\ the ones of the elastic, as well as those of the inelastic fixed window scans, are normalised to the elastic intensity at 2\,K.\\
For the bulk ionic liquid, around 100\,K a drop in the elastic intensity (see Fig.~\ref{fig:FWS_1D}~a) occurs, while that in the $\pm 2\,\text{\textmu eV}$ channels (see Fig.~\ref{fig:FWS_1D}~b) starts to rise considerably. Since bulk [BuPy][Tf$_2$N] is at this temperature still in its crystalline state, this is not the result of any diffusive motion of the molecules as a whole, but originates in the activation of the dynamics of subsets of atoms, like those of methyl groups \cite{Burankova2014} and segmental rotations of the butyl group. When with increasing temperature the dynamics accelerate, the energy transfer between the neutrons and the sample is distributed over a growing energy range and consequently the intensity at the observed $\Delta E = \pm 2\,\text{\textmu eV}$ energy transfer first increases and then decreases. At approx.\ 295\,K the sudden drop/rise in the elastic/inelastic intensity indicates the melting of the bulk ionic liquid.\cite{Noda2001,Tokuda2006}
The situation, however, is completely different in the case of the nanoconfined [BuPy][Tf$_2$N]. The rise of the inelastic intensity at $\Delta E = \pm 2\,\text{\textmu eV}$ from 100\,K on is less steep than in the bulk. But more interestingly, there is no abrupt intensity increase, as for the melting of the bulk liquid, but the intensity starts to gradually rise already at much lower temperatures, such that there are two overlapping peaks. Furthermore, there is a clear pore size dependence. The maximum of the two peaks is successively shifted towards higher temperatures with decreasing nanopore size, while the maximum intensity of the peaks is decreasing. The latter, together with the inverse trend in the elastic intensity, suggests a growing part of immobile ions with respect to their centre-of-mass diffusion.\\
The finding of a gradual melting in the confined versus a discontinuous melting in the bulk state could be related to interfacial melting, i.e.\ the gradual mobilisation of the molecules starting from the carbon pore wall upon heating. Analogous observations have been made for many other liquids in confined geometries \cite{Knorr2008, Schaefer2008, Huber2015}, in particular also at graphitic interfaces \cite{Maruyama1992}. Furthermore, the pore-size dependence of the melting point reduction is a common phenomenon for liquids in porous materials.\cite{Huber1999, Christenson2001, Alba-Simionesco2006, Huber2015}
The Gibbs-Thomson equation, which is strictly valid only for large pores, predicts that the melting point shift scales with the inverse of the pore diameter, where the shift direction depends on the type of interaction of the confined material with the pore wall surface.\cite{Alba-Simionesco2006, Jackson1990} Due to the broad pore-size distribution of the CDCs (see Fig.~\ref{fig:PSD}), it can be expected that this effect significantly contributes to the observed melting transition broadening.

\subsubsection{Low-temperature cation dynamics of the bulk ionic liquid}

\begin{figure}
	\vspace*{4mm}
	\includegraphics{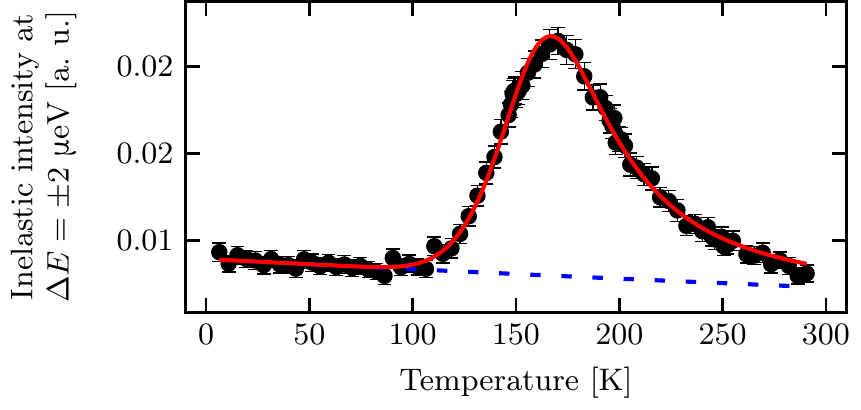}
	\caption{Fit to the elastic fixed window scan data of the low-temperature dynamics of the bulk ionic liquid with the model as delineated in the text.}
	\label{fig:FWS_bulkIL_Fit}
\end{figure}

To further interpret the above findings, first the low-temperature dynamics of the bulk ionic liquid are analysed. For localised motions the incoherent dynamic structure factor can in the simplest case be written as
%
%
\begin{equation}
	S^{\text{local}}(Q,T,\omega) = A(Q) \delta(\omega) + (1-A(Q)) \cdot \frac{1}{\pi} \frac{\gamma_1(Q,T)}{\gamma_1^2(Q,T) + \omega^2} \ \ ,
	\label{eq:FWS_scattering_law_local}
\end{equation}
%
%
where the elastic incoherent structure factor (EISF) $A(Q)$ gives information about the geometry of these motions.\cite{Bee1988} $A(Q)$ is considered here to be temperature-independent. Eq.~\ref{eq:FWS_scattering_law_local} is convoluted with the instrumental resolution and therefore also dynamic processes slower than the resolution of the respective instrument may appear to be elastic. The spectral function is assumed in Eq.~\ref{eq:FWS_scattering_law_local} to be a Lorentzian with the half width at half maximum (HWHM) $\gamma_1$ that is given by the inverse of the relaxation time $\tilde{\tau}^{\text{local}}$ of the corresponding dynamics. This relaxation time is presumed to have an Arrhenius temperature dependence \cite{Grapengeter1987, Frick2012}, i.e.\
\begin{equation}
	\tilde{\tau}^{\text{local}}(T) = \tilde{\tau}_{\infty}^{\text{local}} \cdot \exp\left(\frac{E_{\text{a}}^{\text{local}}}{RT}\right) \ ,
	\label{eq:FWS_Arrhenius_relaxation_time_local}	
\end{equation}
where $R$ is the universal gas constant, $\tilde{\tau}_{\infty}^{\text{local}}$ the high temperature limit of the relaxation time and $E_{\text{a}}^{\text{local}}$ the activation energy of this dynamical process. For the methyl group dynamics a systematic $Q$-dependence of the HWHM is neither expected \cite{Chahid1994}, nor found here (see left panel of Fig.~\ref{fig:IFWS_bulkIL_3D}). Therefore, the intensity, integrated over all for the fixed window scans available $Q$s is analysed. Since furthermore, the expression in Eq.~\ref{eq:FWS_scattering_law_local} is evaluated for $\omega = \pm 2\,\text{\textmu eV} / \hbar =: \omega_{\text{IFWS}}$ only, the temperature-dependent intensity of the fixed window scan, as depicted in Figs.~\ref{fig:FWS_1D} and \ref{fig:FWS_bulkIL_Fit}, is described by
\begin{equation}
	I(T) = e^{-a \cdot T} \cdot \left[I_0 \cdot \frac{1}{\pi} \frac{\tilde{\tau}^{\text{local}}(T)}{1+\left(\tilde{\tau}^{\text{local}}(T)\right)^2 \cdot \omega^2_{\text{IFWS}}} + c_0 \right] \ .
\end{equation}
$I_0$ is an intensity factor and $c_0$ a constant offset. Both are containing a contribution originating in the convolution of the dynamic structure factor (Eq.~\ref{eq:FWS_scattering_law_local}) with the resolution function of the instrument.\cite{Frick2012}
The exponential prefactor describes the temperature-dependence of the Debye-Waller factor \cite{Lechner1991}, determined by a respective fit to the data in the low-temperature region of the corresponding EFWS (cp.\ Fig.~\ref{fig:FWS_1D}~a), before any molecular dynamics aside from lattice vibrations sets in. The constant offset $c_0$ is ascertained in the same temperature region, but from the IFWS.
As can be seen from Fig.~\ref{fig:FWS_bulkIL_Fit}, the employed model fits the measured data very well. It yields an activation energy of 9.7\,kJ/mol for these localised motions. This value is larger than those obtained by \citet{Burankova2014}, employing a three-site jump and a rotational diffusion model, yielding 6.5\,kJ/mol and 7.0\,kJ/mol, respectively, for the analysis of an EFWS at the bulk [BuPy][Tf$_2$N] with a deuterated pyridinum ring.
This is however not surprising, because our cations are fully protonated and thus the determined activation energy is to be considered as an average over all localised dynamics of the molecules. Beneath methyl group dynamics and segmental rotations of the butyl, this includes motions of the pyridinium ring.

\subsubsection{Nanoconfined cation dynamics}

In a next step, the above approach is extended to analyse the molecular dynamics of the ionic liquid confined in the nanoporous carbon over the whole measured temperature range. Contrary to the bulk liquid, the localised dynamics and the ones, ascribed to the translational diffusion of the molecule, can not be analysed separately, since their measurement signals within the IFWS overlap with respect to their temperature range. As a consequence the following ansatz combining both dynamics is chosen to describe the intensity of the IFWS:
\begin{equation}
	I(Q,T,\Delta E) = e^{-a\cdot T} \cdot \left[ I_0(Q) \cdot S^{\text{dyn}}(Q,T,\Delta E) + c_0(Q) \right] \ ,
\end{equation}
where
\begin{equation}
	S^{\text{dyn}}(Q,T,\Delta E) = S^{\text{local}}(T,\Delta E) \otimes S^{\text{global}}(Q,T,\Delta E) \ .
\end{equation}
The dynamic structure factor for the centre-of-mass diffusion of the cation,
\begin{equation}
	S^{\text{global}}(Q,T,\Delta E) = \frac{1}{\pi} \frac{\Gamma_2(Q,T)}{\Gamma_2^2(Q,T) + \Delta E^2} \ ,
	\label{eq:FWS_scattering_law_global_diffusion}
\end{equation}
depends on the wave vector transfer $Q$, while the component considering the localised dynamics, $S^{\text{local}}(T,\Delta E)$, is fully analogous to the one for the bulk ionic liquid (see Eqs.~\ref{eq:FWS_scattering_law_local} and \ref{eq:FWS_Arrhenius_relaxation_time_local}) with a $Q$-independent half width of the spectral function. Because however, the number of free parameters would exceed the number of equations to be solved, further assumptions have to be taken into account: \citet{Embs2012} found in a QENS study of the bulk [BuPy][Tf$_2$N] that the HWHM $\Gamma_2$ of the global component obeys the Singwi-Sj{\"o}lander jump-diffusion model \cite{Singwi1960}, i.e.\
\begin{equation}
	\Gamma_2(Q) = \frac{\hbar D Q^2}{1+D Q^2 \tau_0} \ .
	\label{eq:FWS_Singwi-Sjolander_jump_diff_HWHM}
\end{equation}
In addition, the self-diffusion coefficient $D$ follows an Arrhenius-like temperature dependence \cite{Embs2012},
\begin{equation}
	D(T) = D_\infty \cdot \exp \left( - \frac{E_{\text{a}}^{\text{global}}}{RT} \right) \ ,
	\label{eq:Diffusion_coefficient_Arrhenius}
\end{equation}
and their data suggest, that this also holds for the residence time $\tau_0$, with
\begin{equation}
	\tau_0(T) = \tau_\infty \cdot \exp \left( \frac{E_{\text{a}}^{\text{global}}}{RT} \right) \ .
	\label{eq:relaxation_time_Arrhenius}
\end{equation}
This model is fitted to the 16 $Q$-values of the IFWS data simultaneously with a common set of parameters: the high-temperature limits of the self-diffusion coefficient and residence time, $D_\infty$ and $\tau_\infty$, the activation energies of the localised and diffusive dynamics and the relaxation time of the localised motions. The intensity factor $I_0$ and the EISF $A$ are individual for each $Q$. The exponential prefactor and the background are ascertained as described above for the bulk liquid, but the constant offset $c_0$ is determined for each curve, separately.

\begin{figure}
	\vspace*{4mm}
	\includegraphics{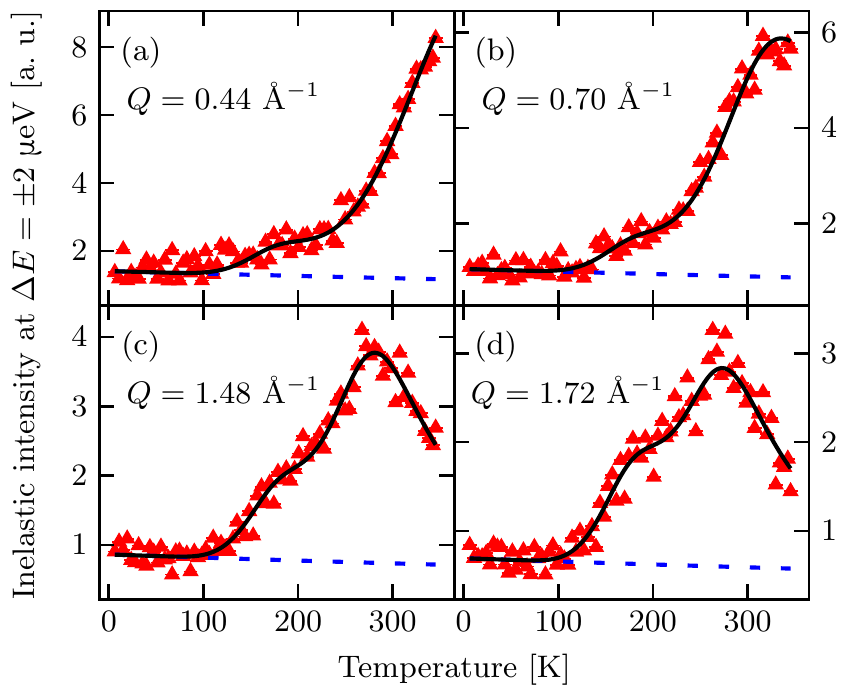}
	\caption{Inelastic fixed window scan at sample {MoC-15+IL} at four exemplary wave vector transfers $Q$ with the corresponding results of the fit to the data, which is performed for all 16~$Q$ simultaneously with a common parameter set and with the model as described in the text.}
	\label{fig:FWS_S4_2D_Fits}
\end{figure}

\begin{figure}
	\vspace*{4mm}
	\includegraphics{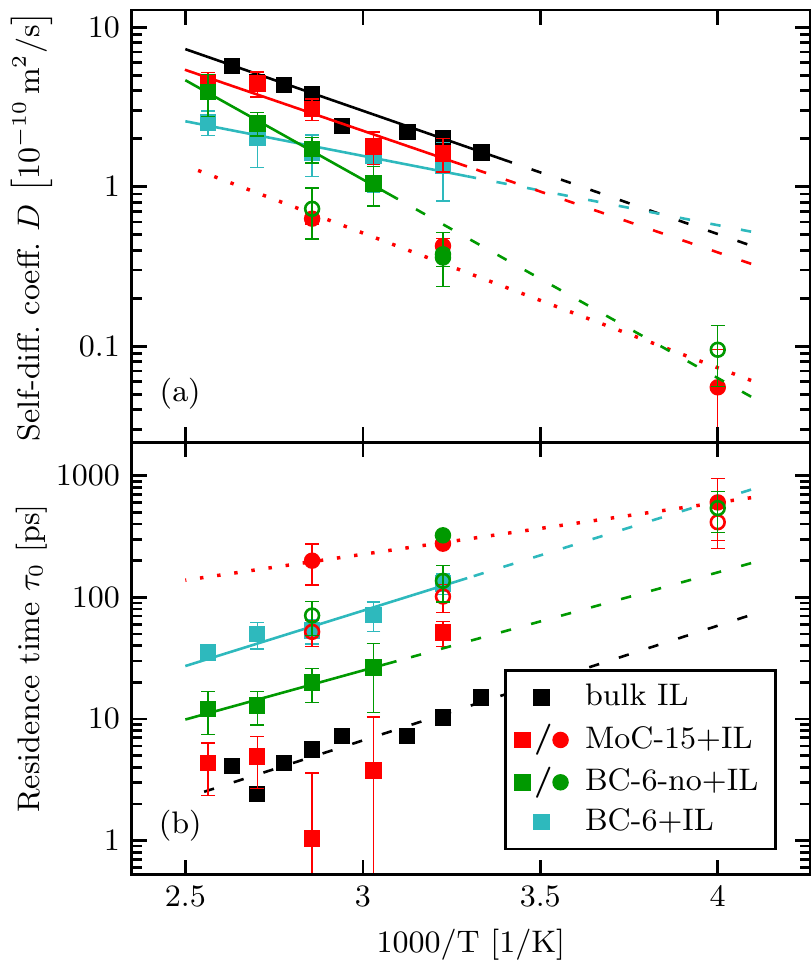}
	\caption{(a) Self-diffusion coefficients $D$ and (b) residence times $\tau_0$ of the [BuPy] cation, as a function of temperature for different nanopore sizes, determined using the Singwi-Sj{\"o}lander jump-diffusion model. Two translational diffusive processes on different time scales are observed. Squares denote data from the \mbox{FOCUS} time-of-flight spectrometer and circles such from the \mbox{IN16B} backscattering spectrometer, at which open circles show values derived from the inelastic fixed window scans. Solid lines are fits with an Arrhenius temperature dependence (see e.g.\ Eq.~\ref{eq:Diffusion_coefficient_Arrhenius}). Dashed/dotted lines are guides for the eye. Bulk data are taken from \citet{Embs2012}.}
	\label{fig:QENS_D_tau0}
\end{figure}

This model reproduces the temperature dependence of the IFWS-intensity very well, as can be exemplary seen in Fig.~\ref{fig:FWS_S4_2D_Fits}, showing the fit to the measurement for sample MoC-15+IL at selected wave vector transfers $Q$.
Table~\ref{tab:IFWS_activiation_energies} lists the activation energies for the translational diffusion of the whole cation, as well as of its localised dynamical processes, obtained from the above analysis.
\begin{table}
 \caption{Activation energies of the localised dynamics and centre-of-mass diffusion of [BuPy] in bulk, as well as in the nanoconfinement of the CDC samples (ordered by decreasing pore size). All data are obtained from the inelastic fixed window scans, except the bulk value of $E_{\text{a}}^{\text{global}}$ (taken from \cite{Burankova2014b, Embs2013, Burankova2017, Embs2012}).}
 \label{tab:IFWS_activiation_energies}
 \begin{ruledtabular}
 \begin{tabular}{lcc}
 & localised dynamics & diffusive motion\\
 & $E_{\text{a}}^{\text{local}}$ [kJ/mol] & $E_{\text{a}}^{\text{global}}$  [kJ/mol] \\
 \hline
 bulk IL 	& 9.7 	& 12.3--14.8\\
 MoC-15+IL 	& 7.1 	& 15.1\\
 BC-6-no+IL	& 6.9	& 14.9\\
 SiC-2+IL	& 6.0	& 12.0
 \end{tabular}
 \end{ruledtabular}
\end{table}
The activation energy of the local motions turns out to be notably smaller in nanoconfinement compared to bulk and it furthermore diminishes with decreasing pore size. For the activation energy of the centre-of-mass diffusion of the cation the same trend is observed: Although the magnitude of the activation energy is for all pores within the range determined for the bulk \cite{Embs2012}, it decreases systematically with the pore size.
The found diminution of the activation energy in the nanopore confinement may appear counterintuitive on the first view, because the maxima of the curves from the IFWS (see Fig.~\ref{fig:FWS_1D}) are successively shifted to higher temperatures with decreasing pore size. But since, the relaxation time $\tilde{\tau}_{\infty}^{\text{local}}$ (see Eq.~\ref{eq:FWS_Arrhenius_relaxation_time_local}) in contrast increases at the same time, the resulting line width $\gamma_1$ decreases nevertheless.
Such a confinement-induced reduction of the methyl group relaxation velocity is also known from substances, like toluene \cite{Moreno2003, Frick2005} and polymers \cite{Frick2005, Schoenhals2005}, where it is ascribed to the interaction of an immobilised surface layer with the confining pore walls. This assumption is supported by the increasing relaxation time with decreasing pore size, entailing an augmentation of the surface-to-volume ratio and thus strengthening the role of surface-interaction-induced effects.
Similar to the relaxation time, also the residence time $\tau_0$ concerning the diffusion of the cation as a whole increases, while the respective self-diffusion coefficient $D$ decreases, with decreasing nanopore size (see open circles in Fig.~\ref{fig:QENS_D_tau0}).
Thus the corresponding line width $\Gamma_2$ (see Eq.~\ref{eq:FWS_Singwi-Sjolander_jump_diff_HWHM}) is smaller for narrower pores and consequently the maximum of the fixed window scan intensity in Fig.~\ref{fig:FWS_1D} is shifted to higher temperatures.
Such a behaviour is intelligible, because the increasing confinement in smaller pores hinders the molecular motions of the cation and thus prolongates the corresponding relaxation times, while slowing down the diffusion. This effect is further intensified by a growing immobile portion of molecules at the pore walls, decreasing the volume of mobile ions, as will be discussed below.

\subsection{\label{sec:full_qens_spectra} Full quasi-elastic spectra}

\begin{figure}
	\vspace*{4mm}
	\includegraphics{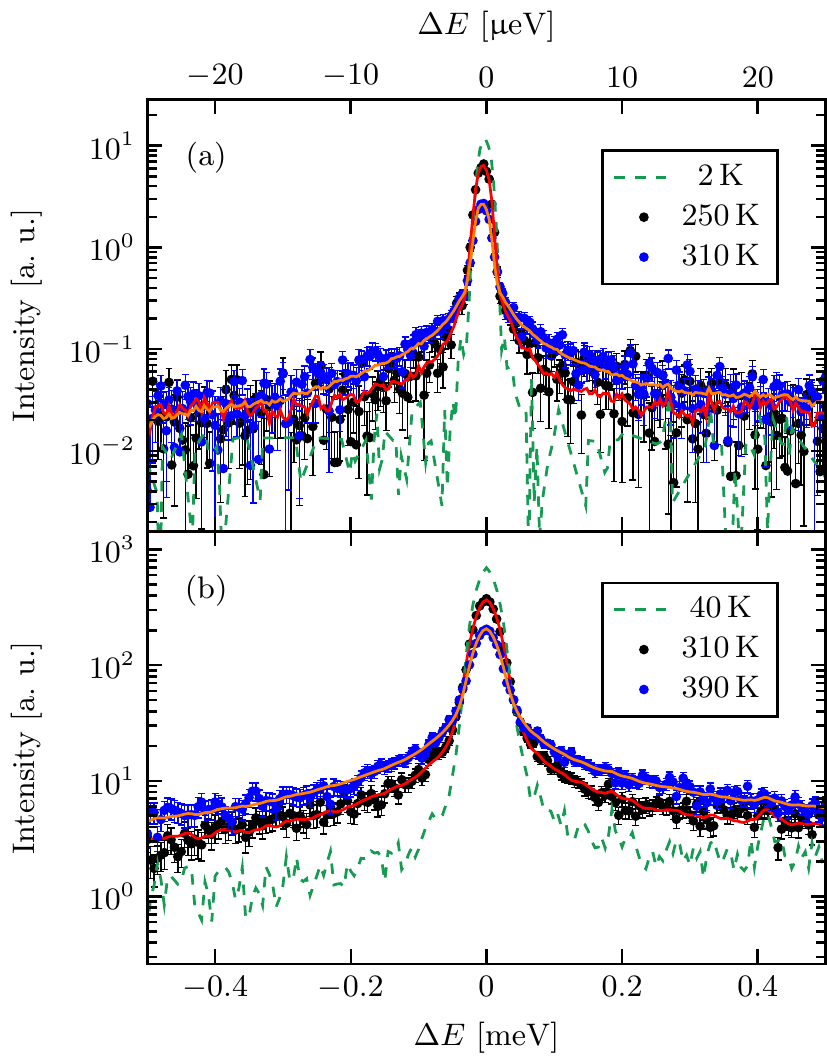}
	\caption{Selected QENS spectra for the ionic-liquid-filled nanoporous carbon MoC-15 as obtained from (a) IN16B ($Q = 1.18\,\text{\AA$^{-1}$}$) and (b) FOCUS ($Q = 1.26\,\text{\AA$^{-1}$}$). Solid lines denote the best fit to the data with the model as described in the text.}
	\label{fig:QENS_spectra}
\end{figure}

\begin{figure}
	\vspace*{4mm}
	\includegraphics{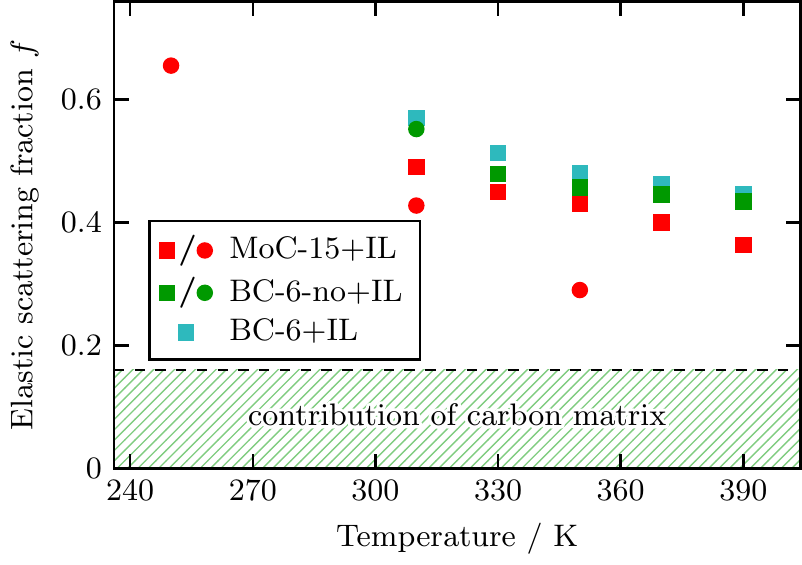}
	\caption{Fraction $f$ of elastic scattering in the quasi-elastic spectra, resulting from immobilized cations of the ionic liquid. Squares denote the fast diffusive process as seen with \mbox{FOCUS} and circles the slower one, observed with \mbox{IN16B}.}
	\label{fig:QENS_elastic_fraction}
\end{figure}

For a more detailed analysis of the cation dynamics of [BuPy][Tf$_2$N] in nanoconfinement, QENS spectra over the full available energy transfer range were acquired at temperatures, selected on the basis of the fixed window scans (see exemplary spectra from both spectrometers in Fig.~\ref{fig:QENS_spectra}).
At \mbox{IN16B} spectra up to 350\,K were measured, while at \mbox{FOCUS} with its lower resolution but wider energy window, a temperature range from 310\,K to 390\,K is covered. For the analysis of the resulting QENS spectra the following dynamic structure factor is applied:
\begin{equation}
	S(Q,\Delta E) = f \cdot \delta(\Delta E) + (1-f) \cdot S^{\text{dyn}}(Q,\Delta E)  \ ,
	\label{eq:QENSspectra_scattering_law_1}
\end{equation}
with
\begin{equation}
	S^{\text{dyn}}(Q,\Delta E) = S_{\text{1}}(Q,\Delta E) \otimes S_{\text{2}}(Q,\Delta E) \ .
	\label{eq:QENSspectra_scattering_law_2}
\end{equation}
To $S(Q,\Delta E)$ a linear background is added and for the fit to the data, it is further convoluted with the resolution function of the instrument, using a measurement of the particular sample at 2\,K or 40\,K in the case of \mbox{IN16B} and \mbox{FOCUS}, respectively.
In the spectra a considerable part of elastically scattered neutrons is observed. Therefore, the fraction~$f$ of immobile scatterers is needed to be incorporated within the scattering law. To keep the model computationally as stable as possible, the elastic fraction is considered for both components of the scattering law, i.e.\ these cations are assumed to exhibit neither translational diffusion, nor localised motions. The scattering laws  $S_{\text{1}}$ and $S_{\text{2}}$ are fully analogue to $S^{\text{local}}$ and $S^{\text{global}}$ used for the analysis of the fixed window scans (see Eqs.~\ref{eq:FWS_scattering_law_local} and \ref{eq:FWS_scattering_law_global_diffusion}).
However, the HWHM $\Gamma_1$ and $\Gamma_2$ of the corresponding Lorentzians are treated as free parameters, now.\\
The line width $\Gamma_1$ of the data acquired at \mbox{FOCUS} is found to be in the range of {60--250\,\textmu eV}, depending on the temperature and shows no clear wave vector dependence. It has a comparable magnitude as in the case of the bulk liquid \cite{Embs2012,Burankova2014b} with a tendency to lower values in the confinement. Furthermore, it exhibits a certain pore size dependence, hinting towards slower dynamics in narrower pores. This component in these QENS spectra is ascribed to localised motions, like the rotation of methyl end groups.\cite{Embs2012} $\Gamma_2$ the HWHM of the second component, however, is found to have a $Q$-dependence according to the Singwi-Sj{\"o}lander jump-diffusion model (see Eq.~\ref{eq:FWS_Singwi-Sjolander_jump_diff_HWHM}). The corresponding self-diffusion coefficients $D$ and residence times $\tau_0$, determined by a fit of the model to the data, are shown as squares in Fig.~\ref{fig:QENS_D_tau0} as a function of temperature for the samples of different pore sizes. For sample SiC-2 unfortunately no reliable information about the self-diffusion dynamics of the ionic liquid could be derived from the QENS spectra.\\
As one can see, the self-diffusion coefficients are notably smaller compared to the bulk values and show a pore size dependence in a way, that with decreasing size of the confinement also $D$ decreases. Alike, the residence time increases with decreasing pore size. Both, the self-diffusion coefficient and the residence time exhibit an Arrhenius-like temperature dependence, but with different activation energies.\\
This is in contrast to the assumptions for the analysis of the IFWS and might explain the differing activation energies obtained from the full spectra, where in deviation from the IFWS results no clear pore-size dependence is found. For the activation energy related to the self-diffusion coefficient of [BuPy][Tf$_2$N] within the nanoconfinement of MoC-15 a value of 15\,kJ/mol is found. For BC-6-no and BC-6 activation energies of 23\,kJ/mol and 9\,kJ/mol are determined, respectively. At first, this appears to be surprising, since according to Fig.~\ref{fig:PSD}, both samples have a very similar distribution of pore widths. However, they underwent different synthesis pathways: BC-6 was air oxidised before the final vacuum annealing step, while BC-6-no was not. The introduction of surface oxide in this intermediate step likely changes the carbon--carbon bonding, such that it is reasonable to assume that this resulted in different pore morphologies and pore walls roughnesses during the vacuum annealing, when surface functional groups --- like oxygen --- are removed. Indeed, there are hints from molecular dynamics simulations that such differences influence the mesoscopic structure and dynamics of an ionic liquid in carbon nanoconfinement.\cite{Monk2011, Vatamanu2011, Seebeck2020}
Also the broad pore-size distribution is presumably of relevance here, because the QENS methods measure the dynamics averaged over all pores. Consequently, also the associated activation energies underlie such an averaging.
While the translational diffusion of the ionic liquid in the pores of MoC-15 has an activation energy similar to the bulk value of 12.3--14.8\,kJ/mol \cite{Burankova2014b, Embs2013, Burankova2017, Embs2012}, the one for BC-6-no is higher, whereas it is smaller for BC-6. It is imaginable that the interaction with the pore wall and structural changes of the ionic liquid that are provoked by the respective pore morphology lead to an alteration of the ionic liquid's dynamics and the related activation energies. Indeed, an activation energy reduction of the translational self-diffusion with respect to the bulk is for example found for hexane in porous silica, where it further decreases with decreasing pore size.\cite{Baumert2002} Alike for supercooled water in silica nanopores such a pore-size dependence of the activation energy is observed.\cite{Liu2006}\\
Similarly in the case of the data acquired on \mbox{IN16B} with its higher resolution but smaller dynamic range compared to \mbox{FOCUS}, two dynamic components are found. One of these is ascribed to the superposition of the slow localised dynamics, as found from the IFWS, and the diffusive motion, as seen with \mbox{FOCUS}. However, since they possess both a comparable HWHM, a separation of both components is challenging and the analysis is focussed to the second, much narrower component, here. The latter is of diffusive nature and its HWHM exhibits again a behaviour, following the Singwi-Sj{\"o}lander jump-diffusion model. The corresponding self-diffusion coefficients obtained from these \mbox{IN16B} spectra are almost one order of magnitude smaller (see filled circles in Fig.~\ref{fig:QENS_D_tau0}~a) than those found with \mbox{FOCUS}, when observing molecular motions at a different time scale. The respective residence time $\tau_0$ on the other hand is increased by one order of magnitude. The self-diffusion coefficients determined from full QENS spectra (more sensitive to spectral shape) are in a very good agreement with those derived from the IFWS (more sensitive to temperature behaviour) above (see open circles in Fig.~\ref{fig:QENS_D_tau0}~a). However, it should be noted that the dynamics inferred from the IFWS are the result of a simplified model containing only one component for each, the centre-of-mass diffusion of the whole cation and the localised dynamics, respectively. Consequently, the derived dynamic properties need to be considered as averaged over the fast and the slow species.\\
Interestingly, we see not only two translational diffusive motions at different time scales, but also a fast and a slow localised dynamic, ascribed amongst others to methyl group motions. It is reasonable to assume that cations exhibiting a sluggish translational diffusion are also those that correspondingly possess slow localised dynamics.
The existence of several translational diffusive motions on different dynamic length and time scales is not unusual for ionic liquids in carbon nanoconfinement \cite{Chathoth2012, Chathoth2013} and even in the bulk state \cite{Ferdeghini2017, Berrod2017, Burankova2018}, where it is often related to a nanoscale structural organisation.\cite{Triolo2007, Russina2012, Russina2017} Also for bulk [BuPy][Tf$_2$N] such a dynamic heterogeneity is found.\cite{Burankova2014b}
In these cases, one dynamic component is regarded as the diffusion of ions bound in ionic aggregates, while the other, long range diffusive motion is considered to take place in the free liquid between these clusters.\cite{Ferdeghini2017, Berrod2017, Burankova2014b, Burankova2018, Ferdeghini2017}
In this connection, a disturbance of the Coulombic charge ordering inside conductive pores, like those of carbon materials, is found e.g.\ by X-ray diffraction experiments and Monte Carlo simulations \cite{Futamura2017, Kondrat2011}, at which this effect is stronger in smaller pores \cite{Futamura2017, Kondrat2014, Kondrat2010}.
Molecular dynamics simulations deliver a further interpretation approach.\cite{Singh2011, Li2013} They find that ions in molecular layers being closer to the pore walls, compared to those in the pore centre, exhibit decreased diffusion coefficients and thus give rise to heterogeneous cation dynamics.\cite{Singh2011, Li2013}\\
At all temperatures the measured quasi-elastic spectra at both spectrometers contain a considerable fraction of elastically scattered neutrons, as already mentioned above. While only a small part of this can be ascribed to the carbon matrix (see estimation \cite{comment_immob} of this contribution in Figs.~\ref{fig:FWS_1D} and \ref{fig:QENS_elastic_fraction}), the predominant portion originates from non-moving cations of [BuPy][Tf$_2$N].
This immobile fraction is decreasing with rising temperature, but even at 390\,K, i.e.\ around 90\,K above the bulk melting point, there is still a notable fraction of immobilised cations. The formation of pore-wall-adsorbed, immobile surface layers is a well-known phenomenon of complex molecules in nanoconfinement \cite{Frick2005, Kusmin2010, Hofmann2012, Huber2015, Richter2019}, as is in the case of ionic liquids in nanoporous carbons.\cite{Chathoth2013, Chathoth2013, Banuelos2014, Dyatkin2018, Dyatkin2018a}
From Fig.~\ref{fig:QENS_elastic_fraction} it appears that there are only few differences between the different specimens concerning the immobile fraction. One of the reasons could be the large volume fraction of micropores, present in all samples, where the cations might be expected to remain largely immobile.
However, there seems to be a weak tendency towards a higher portion in samples with smaller pores, although one should be aware of the uncertainties due to the carbon matrix contribution \cite{comment_immob}. Nevertheless, such a trend is reasonable, because in narrower pores the surface-to-volume ratio is higher and therefore these immobile surface layers occupy a larger fraction of the overall pore volume. The pore-size dependence of the immobile fraction is also supported by the EFWS in Fig.~\ref{fig:FWS_1D}~a. While the elastic intensity of MoC-15+IL with its comparatively large pores appears to saturate at high temperatures only somewhat above the elastic contribution of the carbon matrix, the saturation value of the elastic intensity for the samples with smaller pores is considerably higher and increasing with decreasing pore size. The immobile fraction concerning those molecules exhibiting the slower of the two translational diffusive motions in the nanoconfined ionic liquid appears to be lower than that of the faster diffusing cations (cp.\ circles and squares in Fig.~\ref{fig:QENS_elastic_fraction}). This is because the time scale of the slow dynamics is below the resolution of \mbox{FOCUS} and thus they appear to be immobile there, giving rise to a seemingly elastic contribution to the scattering signal, while they are seen as mobile within the higher resolution of \mbox{IN16B}.

\section{Conclusions}

The cation dynamics of a room-temperature ionic liquid under the nanoconfinement of porous carbons with different pore sizes have been analysed as a function of temperature using quasi-elastic neutron scattering techniques. It is shown in a pioneering manner that the analysis of the fixed window scans gives already a quite comprehensive overview over the different dynamic processes appearing at the nano- and picosecond time scale and about their alteration in confinement, when compared to much more time-consuming full QENS spectroscopic data. A further investigation of the full spectroscopic information finds two diffusive motions of the cation on different dynamic time scales, both slower than in the bulk liquid, while a considerable fraction of molecules stays immobile in the restricting carbon nanopores over the whole temperature range. The obtained self-diffusion coefficients are found to exhibit an Arrhenius-like temperature dependence. But not only the translational diffusion of the whole molecule is influenced by the nanoconfinement. Also the localised dynamics of parts of the cation are successively slowed down with decreasing pore size, while the activation energy of this dynamic process also decreases.\\
Interestingly, our experiments on the thermally excited stochastic motions of the cations unambiguously indicate that the ionic liquid in the nanoporous carbon melts already well below the bulk melting point. Thus, we find no hints of confinement-induced freezing of ionic liquids as recently inferred from slit-pore confinement \cite{Comtet2017}. By contrast we observe the opposite, clear premelting similarly as is known for many conventional liquids \cite{Christenson2001, Alba-Simionesco2006, Huber2015} and this extends the operating temperature range for potential technical applications of such systems, like supercapacitors.\\ 
From a more materials design perspective our study provides mechanistic insights on the slow dynamics of ionic liquids in micropores. It motivates confinement for low-temperature applications, however, it also suggests that hierarchical pore structures, where micropores coexist with mesopores to simultaneously optimize self-diffusive transport and thus functional dynamics with high specific inner surface areas and thus electrical capacities \cite{Liu2017}. In particular, hierarchical pore structures with pore size distributions mimicking bio-inspired Murray materials, hierarchical materials where the pore size distributions is chosen to optimize this tradeoff \cite{Zheng2017}, similarly as established in biological vascularisations, could be able to solve this dichotomy.\\
Finally, we envision some further analysis on the nature of the two diffusive dynamics of the cation. The structural and dynamic investigation of the ionic liquid inside directed nanopores with well defined pore shapes and sizes appear promising. In particular, our findings of distinct thermal activation energies for molecular self-diffusion in nanoporous carbons with similar pore size indicate the importance of the pore morphology and roughness on the molecular mobility, beyond the pure confinement size. This observation motivates a systematic study of these geometrical parameters on the mobility in combination with molecular dynamics simulations \cite{Scheidler2002}.\\
In addition the influence of an applied electrical potential on the molecular mobility of ionic liquids in carbon nanoconfinement needs more attention.\cite{Mahurin2016} Furthermore, mixtures of ionic liquids with solvents may lead to an optimised diffusivity and ionic conductivity \cite{Osti2017a, Thompson2019} and should be systematically investigated concerning their dynamics in carbon nanopores. Also the exploration of the self-diffusion dynamics of aqueous electrolytes confined in nanoporous carbon materials by neutron spectroscopy could be particularly interesting with respect to the impact of ion confinement and desolvation of ions in confinement in the future.\cite{Prehal2017, Prehal2018}

\begin{acknowledgments}
	 We are indebted to Alexei Kornyshev (Imperial College, London) and Yury Gogotsi (Drexel University, Philadelphia) for useful discussions. Furthermore, we thank the working group of Rolf Hempelmann (Saarland University) for providing the ionic liquid. M.B.\ and P.H.\ acknowledge support by the Deutsche Forschungsgemeinschaft within the project ''Property Changes of Multiphasic Fluids by Geometrical Confinement in Advanced Mesoporous Materials'' (Project No.\ 407319385) and the DFG Graduate School GRK 2462 ''Processes in natural and technical Particle-Fluid-Systems (PintPFS)'' (Project No.\ 390794421).
\end{acknowledgments}


\begin{thebibliography}{91}
\bibitem [{com()}]{comment_immob}%
  \BibitemOpen
  \href@noop {} {}\bibinfo {note} {Estimated to be in the range of 0.1-0.2.
  Estimation from the elastic scattering intensity of an empty CDC matrix in
  comparison to that of an ionic-liquid-filled one with the same amount of CDC,
  both measured at 2\,K, where all neutrons are practically elastically
  scattered. Comparison of the calculated scattering cross sections of the
  empty and the ionic-liquid-filled CDC, weighted with the respective particle
  numbers, confirms this finding.}\BibitemShut {Stop}%
\end{thebibliography}
\end{document}